\documentclass[preprint,showpacs,showkeys,amsmath,amssymb,aps,doublespace]{revtex4}
\usepackage[dvips]{graphicx}
\usepackage{setspace}
\usepackage{amsmath}
\usepackage{amsfonts}
\usepackage{amssymb,color}

\usepackage{graphicx,color}
\usepackage{ifpdf}
\usepackage[latin1]{inputenc}
\newcommand{\ii}{\'{\i}}

\begin{document}

\title{Failure of steady state thermodynamics in lattice gases under nonuniform drive
}

\author{
Ronald Dickman\footnote{email: dickman@fisica.ufmg.br}
}
\address{
Departamento de F\'{\i}sica and
National Institute of Science and Technology for Complex Systems,\\
ICEx, Universidade Federal de Minas Gerais, \\
C. P. 702, 30123-970 Belo Horizonte, Minas Gerais - Brazil
}

\date{\today}

\begin{abstract}

To be useful, steady state thermodynamics (SST) must be self-consistent
and have predictive value.
Although consistency of SST was recently verified
for driven lattice gases under global weak exchange, I show here that it
does not predict the coexisting densities in athermal stochastic lattice
gases under a nonuniform drive.  I consider the lattice gas with nearest-neighbor
exclusion on the square lattice, with nearest-neighbor hopping (NNE dynamics), and with
hopping to both nearest and next-nearest neighbors (NNE2 dynamics).  Part of the system is subject to
a drive $D$ that favors hopping along one direction,
while the other part is free of the drive.  Thus the steady
state represents coexistence between two subsystems, one far from equilibrium and the other in equilibrium,
which exchange particles along the interfaces.  The dimensionless chemical potential $\mu^*(\rho,D) = \mu/k_B T$
for the lattice gas with density $\rho$ is readily determined
in studies of a uniform system.  Under the nonuniform drive, however, equating the chemical potentials of the
coexisting subsystems does not yield
the coexisting densities.  The steady state chemical potential
is, moreover, different in the coexisting bulk regions, contrary to the basic principles of thermodynamics.
These results cast serious doubt on the predictive value of SST.
\end{abstract}


\maketitle

\section{Introduction}

A central issue in nonequilibrium physics
is whether thermodynamics can be extended
to systems far from equilibrium
\cite{oono-paniconi,hatano,hayashi,bertin,tomedeoliveira,evans}.
Such a theory would be a macroscopic description
employing a small number of variables, capable of predicting the final
state of a system following removal of some constraint \cite{callen}.
Although the set of variables needed to describe a nonequilibrium system would be somewhat
larger than required for equilibrium, it should not involve
microscopic details.  Near-equilibrium thermodynamics, for example,
includes the fluxes of mass, energy, and other conserved quantities
as relevant variables \cite{onsager,degroot}.

In this context, a natural first step is to develop a thermodynamics of nonequilibrium
steady states (NESS), and to analyze the simplest possible examples exhibiting such states,
for example, driven stochastic lattice gases \cite{KLS,zia,marro,drnne}
or the asymmetric exclusion process \cite{schutz}.
Sasa and Tasaki \cite{sasa2006}, extending the ideas of \cite{oono-paniconi},
proposed a general scheme of steady state thermodynamics (SST), including definitions
of the chemical potential and pressure in NESS, and developed a theoretical analysis of
the driven lattice gas;
numerical implementations in driven systems are discussed in \cite{hayashi,pradhan}.

A central notion in SST is that of {\it coexistence}.  Consider two systems, each in a steady state,
and weakly coupled to one another, so that they may exchange particles and/or energy.
We say that the systems coexist when the net flux of the quantity or quantities they may
exchange is zero.  In equilibrium, coexistence in this sense corresponds
to chemical and/or thermal equilibrium, marked by equality of $\mu/T$ and $T$,
respectively ($\mu$ denotes chemical potential and $T$ temperature).  To construct a SST, we need
to define intensive parameters for NESS, such that the value
of the parameter associated with particle exchange (a dimensionless
chemical potential, $\mu/T$) is the same when two systems coexist with respect to such exchange,
and similarly, an effective temperature, $T$, if the systems coexist with respect to energy exchange.
The definition of intensive parameters for nonequilibrium systems (such as the zero-range process)
possessing an asymptotic factorization property has been discussed in considerable detail by
Bertin et al. \cite{bertin}.

A particularly simple testing ground for SST is
athermal lattice gases, in which
the intensive variable of interest is $\mu^* \equiv \mu/k_B T$,
which I shall call the chemical potential in what follows.  In equilibrium,
$\mu^* = \mu^*(\rho)$, with $\rho$ the particle density.
In a system subject to a nonequilibrium drive $D$, one might hope to define a
function $\mu^*(\rho,D)$ using coexistence between the driven system
and an equilibrium reservoir of known chemical potential.
A recent study \cite{incsst} confirmed that for driven athermal lattice gases with nearest-neighbor (NN)
exclusion, one may
indeed define a function $\mu^*(\rho,D)$ in this manner.
Under global weak exchange, the coexisting densities in a pair of systems
with different values of $D$ are given by the condition of equal chemical potentials, i.e.,
$\mu^*(\rho,D) = \mu^*(\rho',D')$, where $\mu^*$ describes an {\it isolated} system.
{\it Global} exchange means
that any particle in one system may jump to any site in the other (provided, of course,
that the target site and its nearest neighbors are all vacant).
{\it Weak} exchange corresponds to
the limit of the exchange rate $p_r$ tending to zero.

Although the consistency observed in this rather restricted and artificial situation
is encouraging, for SST to be relevant to laboratory conditions, it must be tested
in more realistic settings.
In this work I study coexistence in the NNE lattice gas subject to a {\it nonuniform}
drive.
I consider the lattice gas with NN
exclusion on the square lattice, with two kinds of dynamics: (i) nearest-neighbor hopping (NNE dynamics) and
(ii) hopping to both nearest and next-nearest neighbors (NNE2 dynamics).  Under a drive $D$
that favors hopping along one direction, the system attains a nonequilibrium steady
state (NESS).  When the drive is only applied to half the system we have (in the steady
state) coexistence between a pair of subsystems, one in equilibrium and the other
far from equilibrium, able to
exchange particles along the interfaces separating them.  Let the particle densities
in the driven and undriven regions be $\rho_D$ and $\rho_0$, respectively.
The principal question is whether the condition $\mu^*(\rho_D,D) = \mu^*(\rho_0,0)$
allows us to predict the coexisting densities, in other words, whether SST has
predictive value for this system.

The balance of this paper is organized as follows.  In Sec. II I define the
models and review the relevant properties of the isolated systems.
Section III presents simulation results for the system under a nonuniform drive.
I close in Sec. IV with a
discussion of the implications of the results for steady state thermodynamics.

\section{Lattice gases with nearest-neighbor exclusion}

The lattice gas with nearest-neighbor (NN)
exclusion is a particle model with a pairwise interaction
that is infinite for distances of zero and unity (in units of the lattice constant),
and zero otherwise.  Thus each particle excludes others from occupying
its own site or any of its first neighbors.  Since there is no characteristic energy scale,
the relation between the density and the chemical potential $\mu^* = \mu/k_BT$,
(and similarly, between $p/k_B T$ and $\rho$, where $p$ is pressure),
is independent of temperature.  Such models are termed {\it athermal}.
The model has been studied extensively as a discrete-space version
of the hard-sphere fluid \cite{runnels,gaunt,ree,fernandes}, and is known to
exhibit a continuous (Isinglike) phase transition to sublattice
ordering at a density of $\rho_c \simeq 0.36774$ \cite{guo}.

We define a stochastic, particle-conserving dynamics for
the lattice gas with NN exclusion via particle hopping.
In the simplest case \cite{drnne}, particles are allowed to hop
only to nearest-neighbor sites (the NNE model).  In equilibrium (drive $D=0$), detailed
balance implies that $P({\bf x}) = P(-{\bf x})$, where $P({\bf x})$ denotes
the probability of attempting a particle displacement ${\bf x}$.
In the presence of a drive, the displacement probabilities on the square lattice are:

\begin{equation}
P(\pm {\bf i}) = \frac{1 \pm D}{4}, \;\;\;\;\;\; w(\pm {\bf j}) = \frac{1}{4},
\label{drive}
\end{equation}

\noindent which reduce to the symmetric case for $D=0$.  Evidently,
$D>0$ favors displacements along the $+x$ axis.  Given periodic
boundaries along this direction, $D \neq 0$ represents a nonequilibrium situation,
corresponding to a force that cannot be written as the gradient of a single-valued
potential function.
In the continuous-time stochastic evolution, each particle is equally likely to
be the next to attempt to hop; the hopping direction is chosen according to Eq. (\ref{drive}).
Any particle displacement satisfying the exclusion condition is accepted.

A defect of the hopping dynamics defined above is that it is nonergodic, independent
of the drive \cite{incsst}.
This is remedied by allowing displacements to second as well as nearest neighbors.
For this dynamics, which we denote NNE2, the displacement probabilities are:

\begin{equation}
P(\sigma {\bf i} + \eta {\bf j}) = \frac{1 + \sigma D}{8},
\label{drive}
\end{equation}

\noindent for $\sigma \in \{-1,0,1\}$, and similarly for $\eta$, excluding $\sigma=\eta=0$.
The enhanced set of possible displacements eliminates configurations inaccessible
under NN hopping only.
The phase diagram of the driven NNE2
model was studied some years ago by Szolnoki and Szabo \cite{drnne2}, who showed that
there is a line of Ising-like phase transitions in the $\rho - D$ plane, with
$\rho_c \simeq 0.35$ for $D=1$.

In the lattice gas with NN exclusion, a site is {\it open} if it and all its NNs are vacant.
(Particles can only be inserted at open sites.)  In \cite{incsst} it is shown that
the chemical potential is given by

\begin{equation}
\mu^*(\rho,D) = \ln (\rho/\rho_{op}),
\label{mustar}
\end{equation}

\noindent where $\rho_{op}$ is the average density of open sites over configurations with $n = \rho L^d$ particles.
The above relation follows from coexistence with a particle reservoir, and holds independently of the
value of $D$, and of the nature of the dynamics (NNE or NNE2).
Our definition of $\mu^*$ is equivalent to
the general definition proposed by Sasa and Tasaki \cite{sasa2006}. It is also consistent with the
zeroth law, as verified in \cite{incsst}.

\begin{figure}[!htb]
\includegraphics[clip,angle=0,width=0.8\hsize]{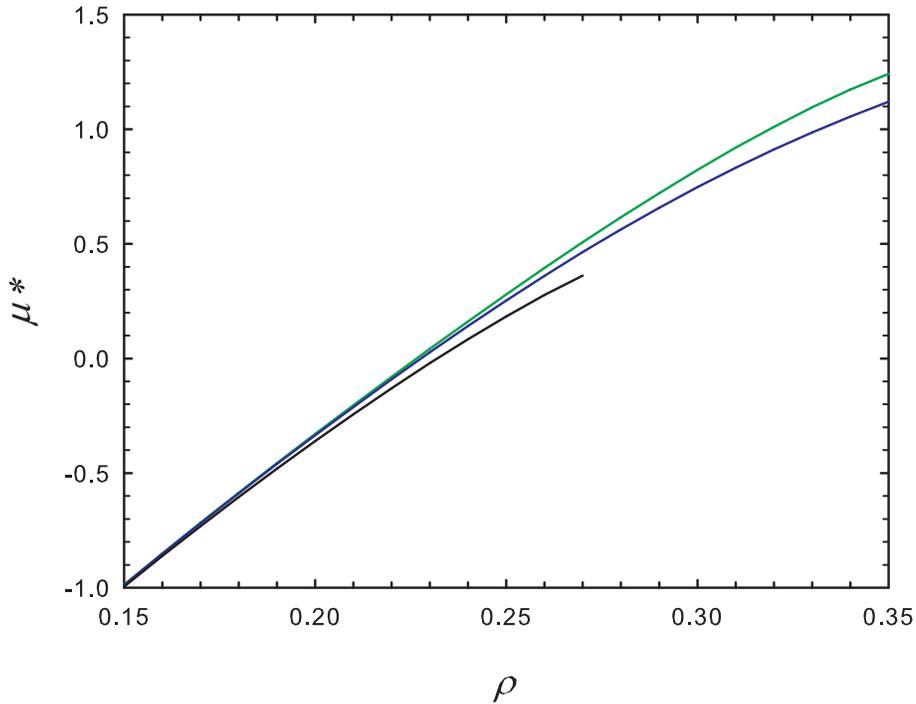}


\caption{\footnotesize{Chemical potential $\mu^*$ versus particle density $\rho$ in the lattice gas with NN exclusion, for
(upper to lower) equilibrium, maximum drive ($D=1$) under nearest- and next-nearest neighbor hopping (NNE2 dynamics),
and maximum drive under nearest-neighbor hopping (NNE dynamics).
Data are simulation results for the square lattice, system size $L=80$.  Uncertainties are smaller than the thickness
of the lines.}}
\label{muall}
\end{figure}

The dependence of $\mu^*$ on
the drive arises because, for a given particle density $\rho$, the density of open sites depends on $D$.
The chemical potential is plotted versus particle density $\rho$ in Fig.~\ref{muall} for
equilibrium, and maximum drive ($D=1$) under NNE and NNE2 dynamics.  Evidently the drive
causes a reduction in $\mu^*$; the reduction is greater in the NNE case.  (Note that the data are limited
to densities smaller than the critical densities of the respective models.)

Studies reported in \cite{incsst} show that if a pair of lattice gases, A and B, driven or not, are allowed to
exchange particles, then in the weak-exchange limit the coexisting particle densities $\rho_A$ and $\rho_B$
are predicted by the condition $\mu^*(\rho_A,D_A) = \mu^*(\rho_B,D_B)$.  This relation holds for both NNE and NNE2
dynamics.
It is important to emphasize that this relation was verified in a highly idealized situation: The contact between
systems is global, in the sense that any site in A can exchange a particle with any site in B.  If one or both of
the systems is driven,
the chemical potentials of the isolated systems are useful in predicting the coexisting densities only
in the limit that the exchange rate $p_r$ tends to zero.  (For finite $p_r$ the chemical potentials
of the coexisting systems are equal, but differ from the values found for the systems in isolation.)
In \cite{incsst} global coupling is used to avoid any possible complication arising from inhomogeneities, such as
localized exchange \cite{pradhan}.  The weak exchange limit is necessary because the strength
of the nonequilibrium perturbation varies with $p_r$.

In the present study I take a single NN exclusion lattice gas on a lattice of $L \times L$ sites
and divide it in two by applying a nonzero drive in half of the system, that is for sites $(i,j)$
with $\frac{L}{2} + 1 \leq j \leq L$.  Thus the boundaries between the driven and undriven regions
(between $j = \frac{L}{2}$ and $\frac{L}{2} +1$, and between $j=L$ and $j=1$), are parallel to the
preferred ($x$) direction.  In the following section I report simulation results for the
stationary density and chemical potential profiles ($\rho(j)$ and $\mu^*(j)$, respectively, the latter
obtained via Eq. (\ref{mustar}), with the $j$-dependent values of $\rho$ and $\rho_{op}$).  If SST
functions correctly, the coexisting bulk densities $\rho_0$ and $\rho_D$ should be given by the
equal chemical potential condition (implying $\rho_D > \rho_0$), and the chemical potential profile should be
flat, $\mu^*(j) = \mu^*(\rho_0,0) = \mu^*(\rho_D,D)$.

\section{Simulation results}

\subsection{NNE dynamics}

I study the NNE lattice gas on square lattices of $L \times L$ sites (with periodic boundaries),
with $L$ ranging from 200 to 800.  Initially, particles are distributed uniformly over the lattice,
by inserting particles at randomly chosen open sites.
For the densities and system sizes considered here, the system
attains a stationary state well before $5 \times 10^6$ times steps, the time allowed for relaxation.
(A time step corresponds to $n$ attempted particle moves.)  Averages are calculated over an additional
set of $5 \times 10^6$-$10^7$ steps, following relaxation.

\begin{figure}[!htb]
\includegraphics[clip,angle=0,width=0.9\hsize]{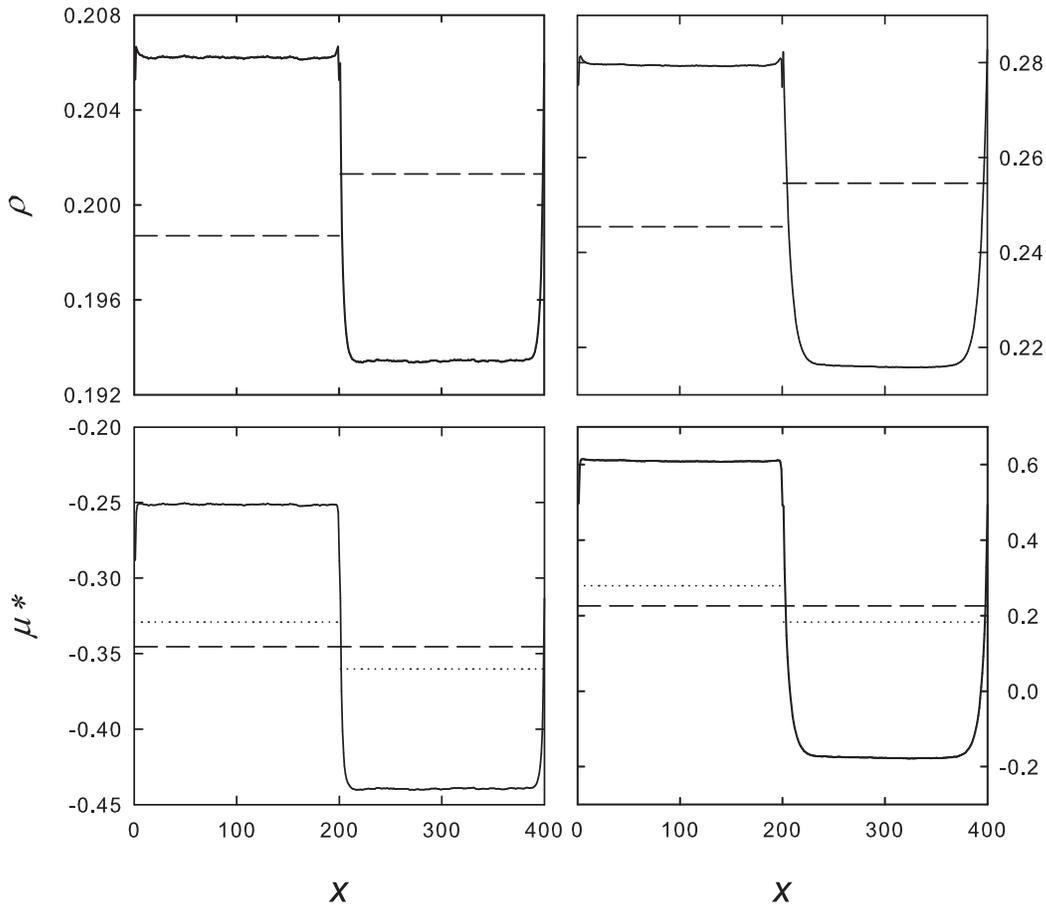}


\caption{\footnotesize{Density (upper) and chemical potential (lower) profiles in the half-driven NNE model.
Drive $D=0$ for $x \leq 200$; $D=1$ for $x > 200$.  Global densities $\overline{\rho} =0.2$ (left) and 0.25 (right).
In the upper panels the dashed lines show the coexisting densities predicted by equating chemical
potentials.  In the lower panels the dashed lines show the expected uniform value of the chemical
potential, and the dotted lines show the chemical potentials for the isolated systems, each at density $\overline{\rho}$.}}
\label{profs}
\end{figure}

\begin{figure}[!htb]
\includegraphics[clip,angle=0,width=0.8\hsize]{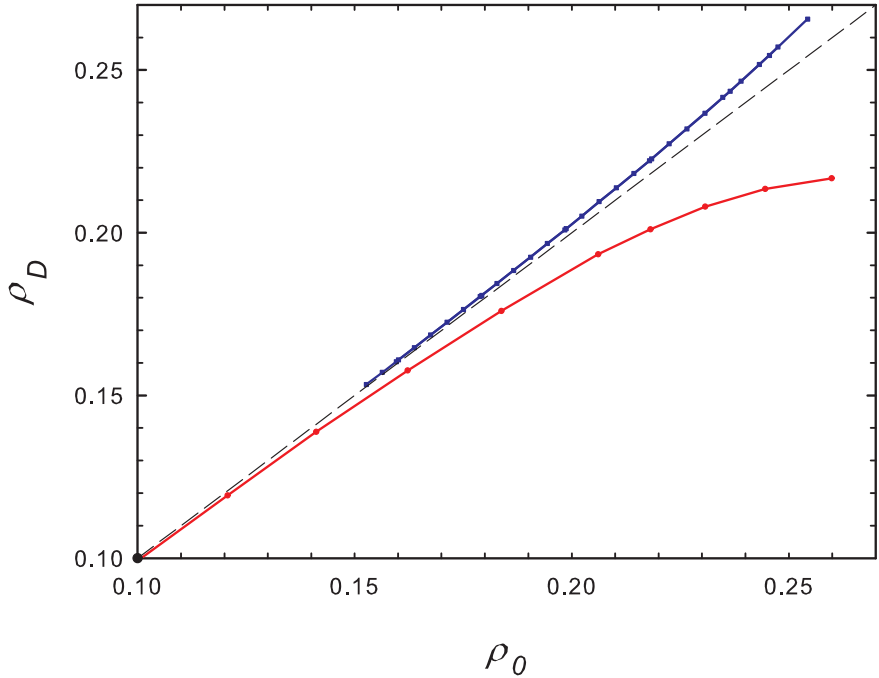}


\caption{\footnotesize{(Color online) Coexisting bulk particle densities $\rho_D$ and $\rho_0$,
in driven and undriven regions, respectively,
in the NNE model, as predicted by equating the chemical potentials in isolated systems (upper curve)
and observed in simulations of the half driven system (lower curve).
Error bars smaller than symbols.
For purposes of comparison, the dashed line represents $\rho_D = \rho_0$.}}
\label{cxnne}
\end{figure}

The stationary properties of the half-driven system are hardly what one would expect based on SST.
Typical density and chemical potential profiles are shown in Fig.~\ref{profs}.
The density is higher in the undriven
region, contrary to the prediction obtained equating the chemical potentials of driven and undriven systems.
The chemical potential profile is clearly nonuniform, more so,
in fact, than if the particle density were uniform at the global density.  Particles have migrated
so as to increase severalfold the difference in chemical potential between the two
regions, rather than diminish it!  Varying the global density, the bulk densities observed in
simulation consistently violate
the expected tendency, that is, $\rho_D < \rho_0$; the difference grows with global density,
as shown in Fig.~\ref{cxnne}.

The density and chemical potential profiles of Fig.~\ref{profs} show well defined bulk regions,
justifying the interpretation of coexisting phases.  The bulk density and chemical potential values, $(\rho_0,\mu^*_0)$ and
$(\rho_D,\mu^*_D)$, fall on the corresponding curves, $\mu^*(\rho,D=0)$ and $\mu^*(\rho,D=1)$,
characterizing the isolated systems.  In other words, the driven and undriven regions
retain their respective bulk properties in the presence of a nonuniform drive.
I note that the chemical potential profile within the nondriven region is essentially flat,
despite significant variations in density, as one would expect in an equilibrium system such
as a fluid confined between repulsive walls.  In the driven region, by contrast, the chemical
potential profile varies over a substantial region ($\sim$10-20 lattice spacings) near the boundaries.
In the region near the boundary, the chemical potential is well approximated by an exponential,
$\mu^*(x) - \mu^*_D \sim e^{-x/\lambda}$, where $x$ measures the distance from the boundary.
The ``healing length" $\lambda$ grows with density;
for a global density of 0.25, I find $\lambda \simeq 5.5$.

Particles migrate to the undriven region, parallel to the chemical
potential gradient (i.e., {\it contrary} to Fick's law),
for all system sizes examined, as well as for a smaller drive ($D=0.5$).
Migration to regions of zero or weaker drive was noted (for low densities) in \cite{drnnend}, in which
the NNE model is subject to a drive which varies linearly with position $j$ in the direction perpendicular
to the drive.  In \cite{ladder} a NNE model on a two-lane ring (one driven, the other undriven)
was found to exhibit
particle migration to the undriven region for smaller global densities, and migration in the
opposite sense for $\rho > 0.3$.  Although a full explanation of the transport mechanism
is not available, it appears \cite{ladder} that diagonal
strings of particles near the boundary between driven and undriven regions favor the transfer of
particles from the former to the latter.  Since the present study is concerned with macroscopic behavior,
we defer further analysis of this question to future work.

\begin{figure}[!htb]
\includegraphics[clip,angle=0,width=0.8\hsize]{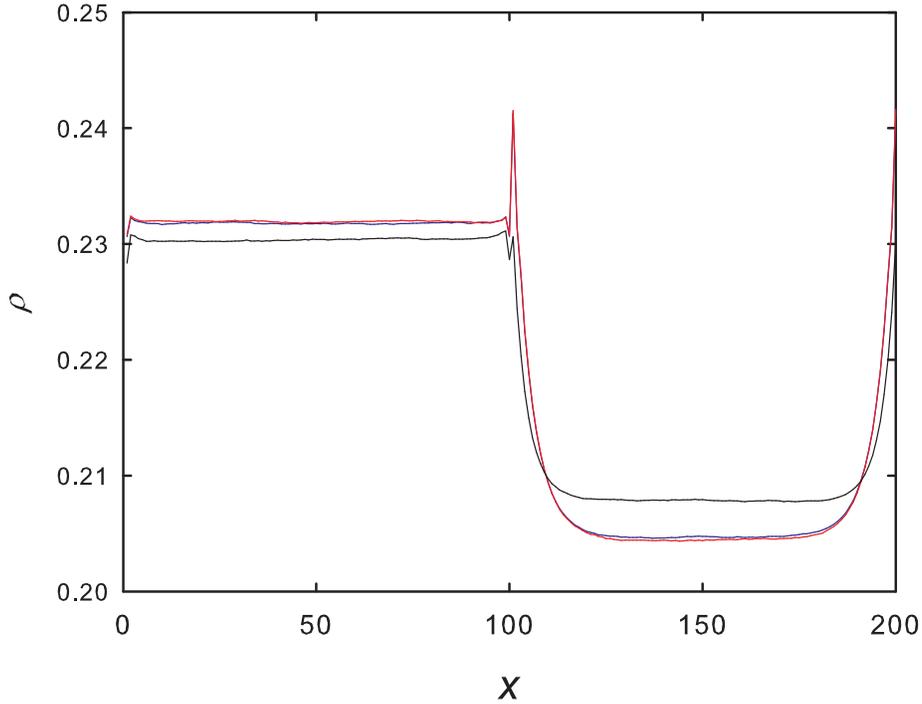}


\caption{\footnotesize{(Color online) Density profiles in the half-driven NNE model with global density 0.22;
drive $D=0$ for $x \leq 100$; $D=1$ for $x > 100$.  Black: $p_r=1$; blue: $p_r=0.005$; red: $p_r = 0.002$.}}
\label{pr200}
\end{figure}

From the macroscopic viewpoint, the driven and undriven regions correspond to systems coexisting under
particle exchange.  One therefore expects the stationary properties to be predicted by SST, which it evidently
does not.  Is the failure due to limited system size?  Studies using $L=200$, 300, 400 and 800
yield essentially the same results, eliminating size as a possible explanation.
Another possibility is that the exchange between the
two regions must be weaker for SST to function, since, as noted in \cite{incsst}, full agreement with the predictions of
SST requires that we take the weak-exchange limit.  Studies in which the
acceptance probability for transfers between the two regions, $p_r$, is small, reveal essentially the same
pattern as observed for $p_r=1$.  In the example shown in Fig.~\ref{pr200}, for global density 0.22,
the departure from the expected behavior is in fact somewhat greater for small values of $p_r$.


\subsection{NNE2 dynamics}

I study the half-driven model with NNE2 hopping dynamics using
simulation parameters similar to those used in the NNE case.
Since the phase transition to sublattice ordering occurs for
$\rho \simeq 0.36$, the studies can be extended to higher
densities than for NNE dynamics.
Studies with $L=200$ and 400 yield the same bulk values for $\rho$ and $\mu^*$, to within uncertainty.

\begin{figure}[!htb]
\includegraphics[clip,angle=0,width=0.6\hsize]{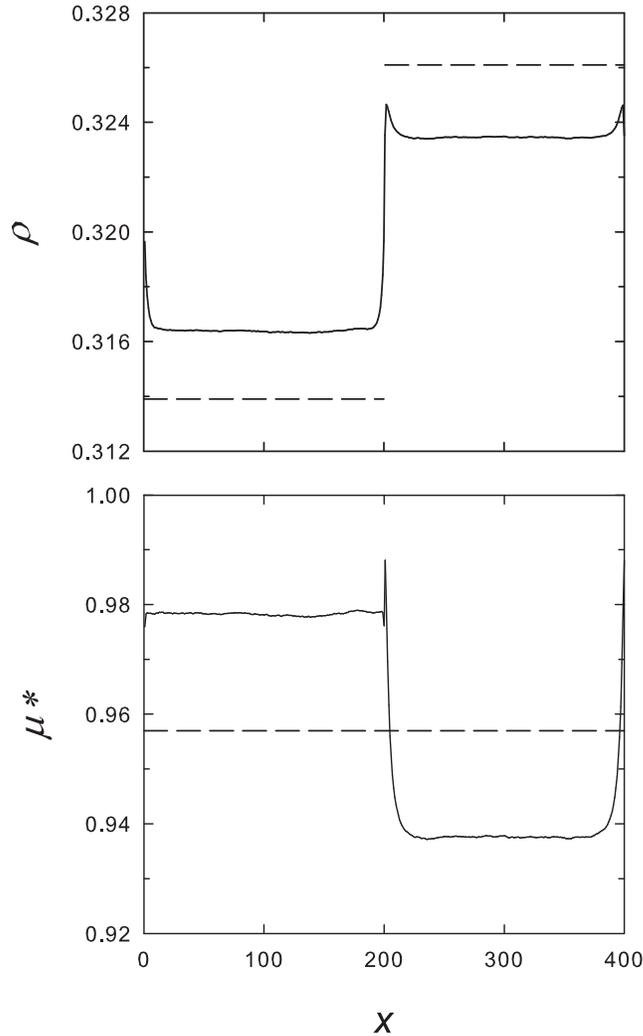}


\caption{\footnotesize{Density (upper) and chemical potential (lower) profiles in the half-driven NNE2 model with
global density $\overline{\rho} =0.32$.
Drive $D=0$ for $x \leq 200$; $D=1$ for $x > 200$.
In the upper panel the dashed lines show the coexisting densities predicted by equating chemical
potentials.  In the lower panel the dashed line shows the expected uniform value of the chemical
potential.}}
\label{prnne2}
\end{figure}

\begin{figure}[!htb]
\includegraphics[clip,angle=0,width=0.8\hsize]{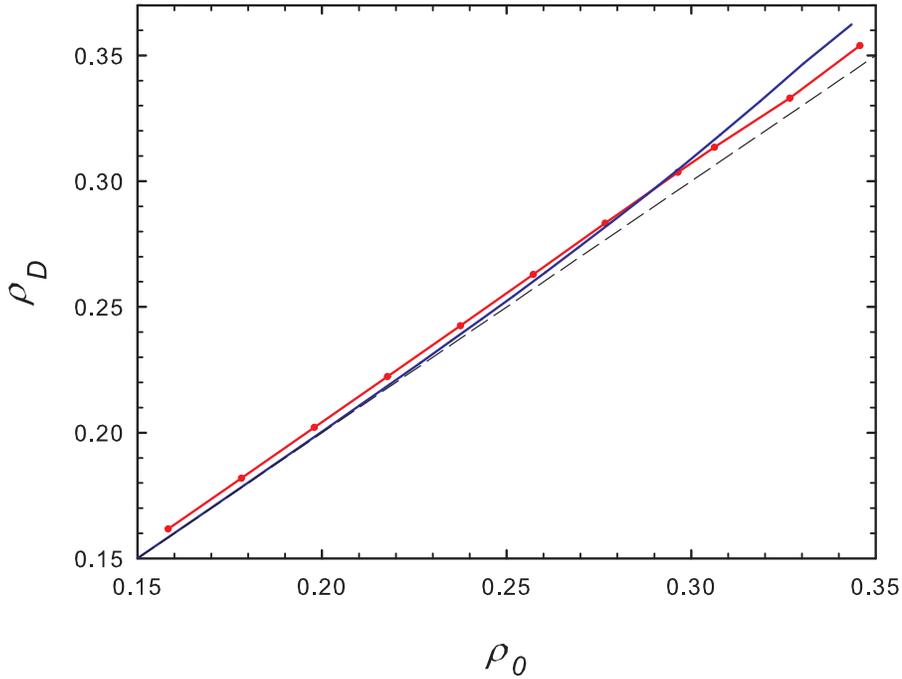}


\caption{\footnotesize{(Color online) Coexisting bulk particle densities $\rho_D$ and $\rho_0$,
in driven and undriven regions, respectively,
of the NNE2 model, as predicted by equating the chemical potentials in isolated systems (lower curve at left)
and observed in simulation (upper curve at left).  For purposes of comparison,
the dashed line represents $\rho_D = \rho_0$.}}
\label{cxnne2}
\end{figure}

\begin{figure}[!htb]
\includegraphics[clip,angle=0,width=0.8\hsize]{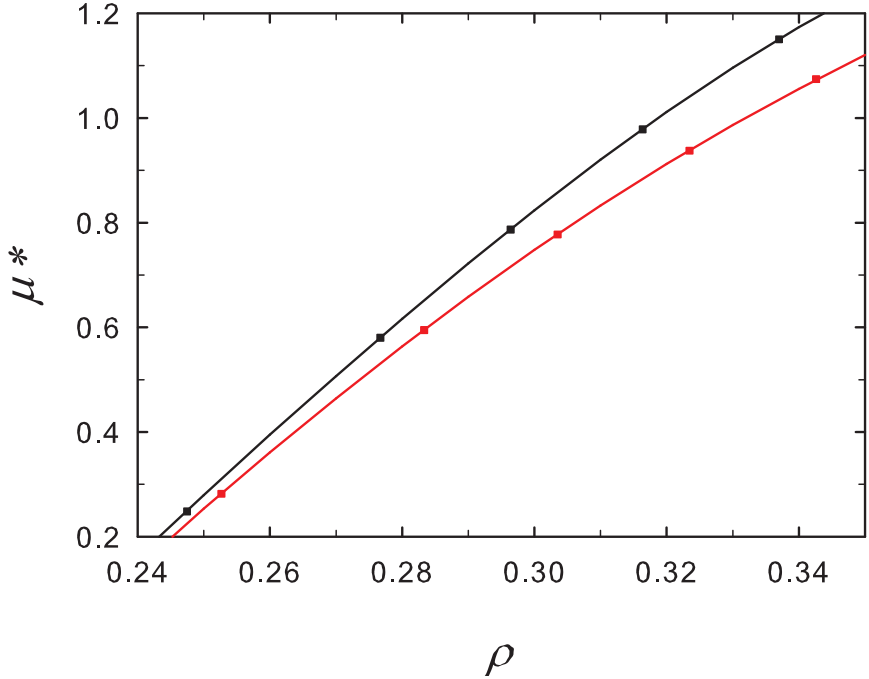}


\caption{\footnotesize{(Color online) Points: bulk chemical potential versus bulk density in driven (lower) and
undriven (upper) regions of half-driven NNE2 model.  The curves show the corresponding values for
isolated systems.  Uncertainties smaller than symbols.}}
\label{murhhd2}
\end{figure}

In contrast to the NNE case, under NNE2 dynamics the stationary density
is higher in the driven region, as predicted by SST.
The observed density and chemical potential profiles are not, however, consistent with those
expected on the basis of SST.  As shown in Fig.~\ref{prnne2}, the bulk densities do not take their expected
values, and the chemical potential profile is not uniform.  Once again, the principal of
equal chemical potentials in systems that coexist under particle exchange is violated.  The general lack of
agreement is evident in the comparison (Fig.~\ref{cxnne2}) between predicted and observed densities
in the driven and undriven regions.  At lower global densities, $\rho_D$ exceeds the value predicted
using SST, while at higher global densities (above about 0.29) the trend reverses.  (The difference
$\Delta \rho = \rho_D - \rho_0 \propto \rho^2$ as the global density $\rho$ tends to zero.)
As in the NNE case, reducing the
acceptance probability, $p_r$, for transfers between the driven and undriven regions only serves to
enhance (slightly) the discrepancy between simulation and SST.
Once again, the bulk values of $\rho$ and $\mu^*$ in the driven and undriven regions agree with those
found for the corresponding isolated systems, as shown in Fig.~\ref{murhhd2}.

\begin{figure}[!htb]
\includegraphics[clip,angle=0,width=0.7\hsize]{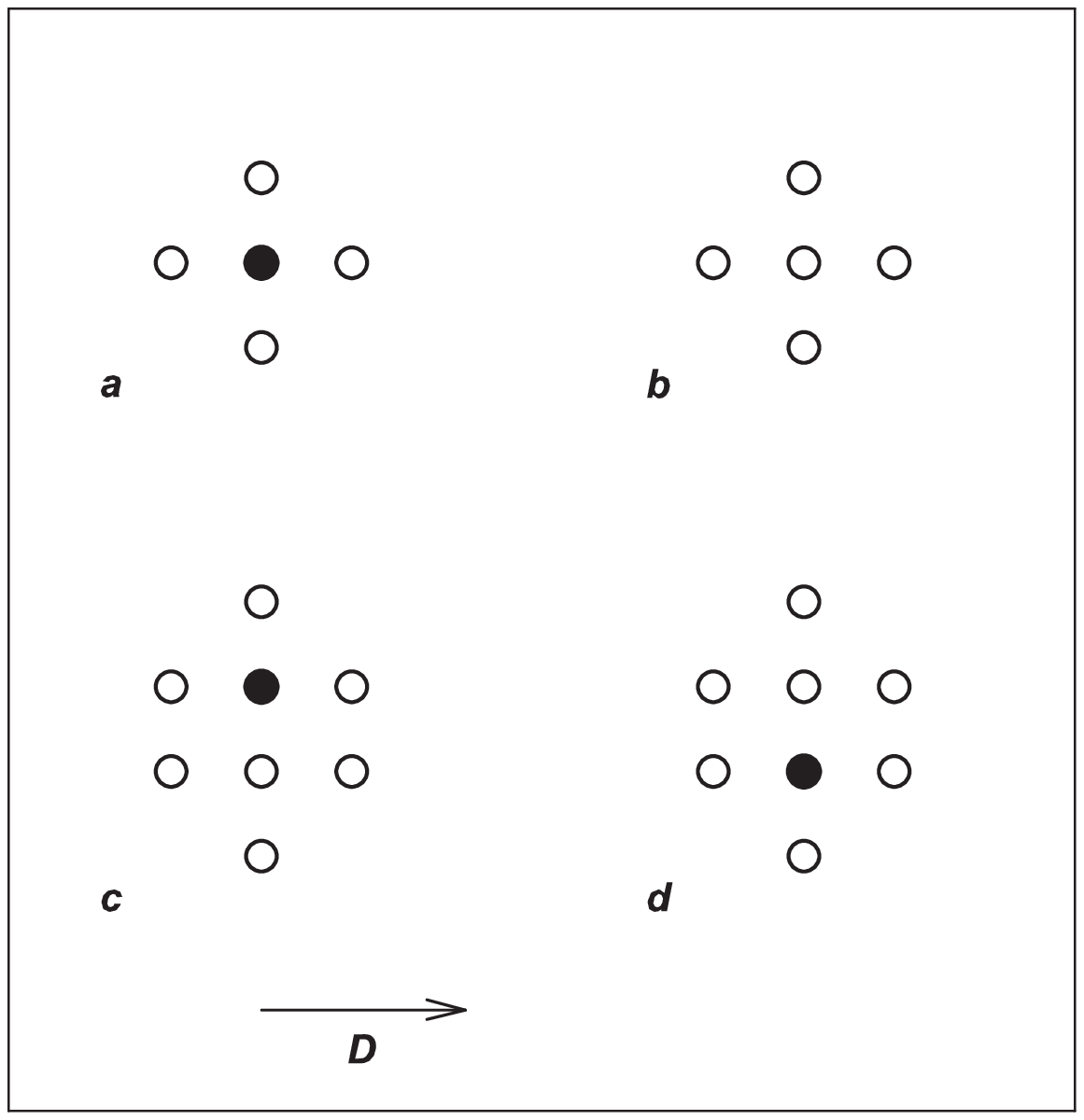}

\vspace{-3cm}

\caption{\footnotesize{Clusters on the square lattice.  A uniform chemical potential requires that the ratio of the probabilities
of clusters $a$ and $b$ be the same everywhere.  A time-independent density profile perpendicular to the
drive {\bf D} requires that at each row parallel to {\bf D}, the probabilities of clusters $c$ and $d$ be equal.}}
\label{clusters}
\end{figure}

\section{Conclusions}

I test the predictive value of steady state
thermodynamics in the simple context of athermal lattice gases under a nonuniform drive.
SST is found ineffective in predicting the coexisting particle densities of
the driven and undriven systems when they exchange particles across the interface,
for both nearest-neighbor hopping (NNE dynamics) and nearest- and
next-nearest hopping (NNE2 dynamics), regardless of the transfer probability $p_r$
at the boundary.  It should be noted that the essential difference between
the setup of Ref.~\cite{incsst}, in which coexistence is attended by equality
of the chemical potential, and the present study, in which it is not, is the mode
of particle transfer (global versus local).
The global transfer scheme of Ref.~\cite{incsst}, while useful for eliminating
complicating effects of inhomogeneities, is not realizable in the laboratory.
Here, by contrast, we study the rather natural situation of a spatially nonuniform drive,
with exchange restricted to the region of contact between the subsystems, as would occur
in practice.
The violation of SST
is generally smaller under NNE2 dynamics, which features a longer
range of particle motion.  This is consistent with the
observation that, extending the range of motion to include all sites, one would have
{\it global} exchange, for which SST is in fact valid.

It is worth contrasting
the condition of (I) thermodynamic coexistence (spatially uniform chemical potential) with that (II)
of a stationary density profile perpendicular to the drive.
On the square lattice, condition I implies that the ratio of the probabilities of the two five-site
clusters ($a$ and $b$) shown in Fig.~\ref{clusters} be independent of position, whereas
condition II simply requires that at each row $j$ along the drive, the
probabilities of clusters $c$ and $d$ be the same.  The condition of uniform $\mu^*$
derives from the definition of chemical potential via exchange with a particle
reservoir \cite{incsst}.  But since the driven lattice gas is not in contact with
such a reservoir, there is no reason for condition I to apply in the presence of a nonuniform
drive.  Condition II does apply (else there would be a current perpendicular to the
drive); it can be satisfied in a multitude of ways, depending on the details of
particle fluxes in each row along the drive.

The failure of SST in this simple
context strongly suggests that it cannot predict the stationary properties of
coexisting nonequilibrium systems, except in the highly artificial case of
weak global exchange studied in \cite{incsst}, and, therefore
that the thermodynamics of nonequilibrium steady states
has rather limited utility.  Tests of SST in other model systems are planned for future work.


\newpage
\noindent {\bf Acknowledgments}

I thank Fabr\ii cio Potiguar for helpful comments.
This work was supported by CNPq and CAPES, Brazil.
\vspace{2em}


\end{document}